# Study of SPIRAL 2 High Radiofrequency Cooler: cooling of very high intensity ion beams

**Ramzi Boussaid**[a*] **,Gilles Ban** [a] **and Jean Francois Cam** [a]

[a] *LPC-IN2P3, ENSICAEN, 6 Boul. Maréchal Juin, 14050 Caen, France.*

*E-mail*: boussaidramzii@gmail.com

Abstract:

In the framework of the DESIR/SPIRAL-2 project, an RFQ Cooler named SHIRaC has been studied. SHIRaC is a key device of SPIRAL-2 to enhance the beam quality required by DESIR. The preliminary study and development of this device has been carried on at LPC Caen (France).The experimental studies of SHIRaC prototype are the goal of this paper. The main addendum of this cooler is the handling and the cooling of beam currents going up to 1 µA ion beams which have never yet been done.

Much effort has been made lately into these studies to develop the appropriate optics, vacuum and RF systems which allow to cool beams of large emittance (~ 80 $\pi$ mm mrad) and high current. The dependencies of SHIRaC's transmission efficiency and the cooled beam parameters in terms of the geometrical transverse emittance and the longitudinal energy spread will be discussed. We will also investigate the beam purity at the optimum cooling conditions.

Results indicate that an emittance reduction of less than 2.5 $\pi$ mm mrad and longitudinal energy spread reduction of more than 4 eV are obtained with more than 70 % ion transmission. The emittance is at the expected values but the energy spread is not.

*Ramzi Boussaid, boussaidramzii@gmail.com

# Contents

# 1. Introduction

In the context of new generation nuclear facilities based on the ISOL method [1] for the production of rare and exotic ion beams with intensities up to 1μA and emittances up to 80 π mm mrad [2], SPIRAL-2 project is installing at GANIL laboratory in France [3, 4, 5]. Such ion beams produced by SPIRAL-2 will subsequently suffer from isobaric contaminations [6].
The DESIR (Désintégration, Excitation et Stockage d'Ions Radioactifs) facility is a low energy beam experiment. It is meant to receive beams from SPIRAL-2 [7, 8]. The high degree of purity required to push experiments towards the limits of stability will be achieved by the implementation in the SPIRAL-2 production building of a HRS (High Resolution Separator) [9, 10].

In order to make the best use of produced beams, the HRS should receive beams of low transverse emittance ≤ 3 π mm mrad and low energy spread ~1 eV. There are now several instruments operating many of the world's radioactive beam installations capable of providing low beam emittances with good efficiency. All are based on a combination of a radiofrequency quadrupole (RFQ) filled with a buffer gas, and an axial field to guide the beam. For that, a buffer-gas radiofrequency quadrupole cooler (RFQC) should be installed in front of the HRS.

The RFQCs are already used in several projects such as LEBIT at NSCL [11], ISCool at ISOLDE [12, 13], TITAN at TRIUMF[14], …etc. Unlike previous RFQCs, which can only handle beams with low intensities ~100 nA and small emittances ~10 π.mm.mrad, the RFQC required by SPIRAL-2 should be able to capture higher intensities and larger emittance of the incoming beam. In order to fully exploit these intensity increases, a new RFQC prototype named SHIRaC (Spiral-2 High Intensity Radiofrequency Cooler) is developed and tested at LPC-Caen laboratory at France. This prototype is devoted both to provide the desirable quality beam for the HRS and to transmit more than 60 % of incoming ions toward the HRS.

SHIRaC gets over several challenges to handle SPIRAL-2 beams thanks to their three systems: an appropriate optics system to capture large emittances, an RF system providing highest RF voltage amplitude to overcome the space charge effects due to the high beam currents [15] and a specific vacuum system to control the buffer gas pressure.

Numerical simulations allowing to optimize the SHIRaC's design and to determine the needed cooling parameters in terms of buffer-gas pressure, RF parameters, electrode voltages…etc are already published [16].

The developments and tests of SHIRaC's systems are presented in this paper. Results of experimental investigations of the Cs ion beam cooling with beam currents up to 1 µA are also reported. The dependencies of the ion transmission and cooled beam properties (which stand for the beam purity, longitudinal energy spread and geometric transverse emittance) on the space charge effects, buffer gas pressure and RF voltage amplitude are outlined. We also investigate the beam purity at the cooling optimum conditions.

## 2. Experimental setup

RFQCs currently play an important role in purifying and preparing radioactive ion beam in today's installations. Their design is a complex process including several physics and technical topics. The research reported here should help to determine the needed HRS specifications, as previously mentioned. In the SPIRAL-2 context, SHIRaC consists of three systems: the optics system, the vacuum system and the electronic system; and its design depends mainly on the beam properties both before and after the RFQ.

The design process of such a device is divided into three main phases:

- The ideal design, to perfectly match with experiments, should provide an RFQC able to handle highest beam current at the entrance of the RFQ and provide the least possible beam quality at the exit.
- The design of the optics system influences the vacuum system running. A compromise has to be found between the necessity of increasing as much as possible the acceptance of the device and having the lower pressure outside the RFQC.
- The design of the electronic system is related to both the vacuum and optics systems. The maximum of the RF voltage parameters (the amplitude and the frequency) are limited by the electrical breakdown phenomenon which depends on the buffer gas environment pressure and the distances between the electrodes.

Taking into account these constraints the RFQC design was developed. The optimum design of the optics, vacuum and electronic systems will now be presented.

## 2.1. SHIRaC optics

The main goal of the SHIRaC project is to improve the SPIRAL-2 beam quality, so a lot of care has to be taken in the design of the optics system.

Similar to the many coolers now used all over the world such as those presented in these references [11, 17, 18, 19], SHIRaC can be divided concerning the optical structure, as seen in figure 1, in three parts: injection and deceleration part, RFQ part and extraction and reacceleration part. The optical design of all the parts was optimized using the software SIMION-3D V8.0 [21].

The injection part contains the electrodes to decelerate the ions down to the input energy into the RFQ. This energy should fall within the expected range 100-200 eV [17], in the present case it is around 140 eV. The beam is injected into the RFQ chamber through the injection plate. A lens of three electrodes is mounted to avoid the ions losses during the deceleration process and to improve the injection of ions.

The RFQ part is the main RFQC chamber. It encloses the buffer gas and the RF quadrupole (RFQ). This part is devoted to confining efficiently the injected beam and to cooling it progressively. To adapt to SPIRAL2/DESIR authentic beams, the internal radius was $r_0$=5 mm. The electrodes' RFQ are segmented to 18 equidistant segments of 40 mm of length. This segmentation allows creating a longitudinal electric field to guide the cooled ions which are then released from the RFQ limit through the extraction plate.

The extraction and reacceleration part contains the electrodes that make the ion beam extract and then accelerate the cooled ions up to their initial energy.

Details of the design and all electrodes dimensions are described in references [16, 20].

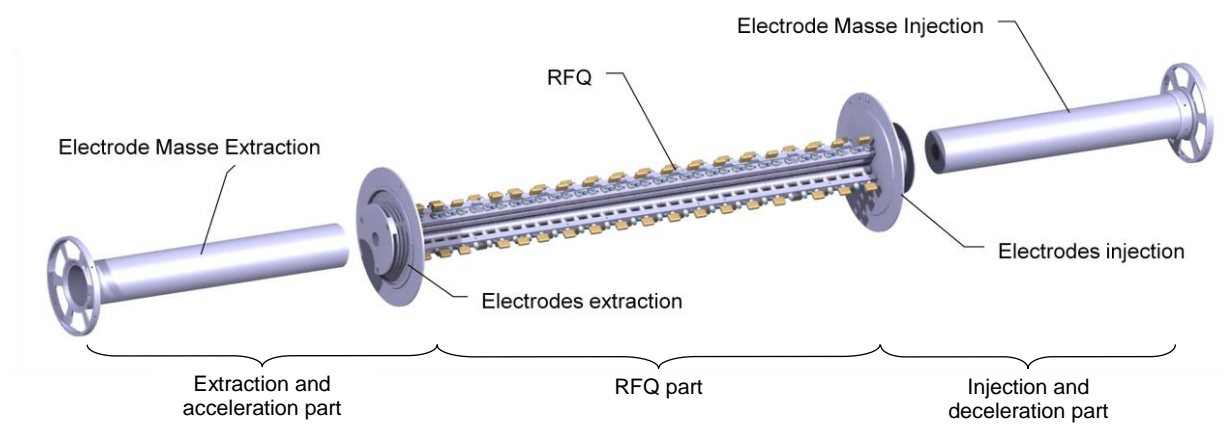

**Figure 1**: Schematic of the SHIRaC optics: design of three parts of SHIRaC

## 2.2. Vacuum system

To run an RFQC we must use a buffer gas inside a RFQ chamber that helps to cool the ion beam. The cooling is provided subsequent to the successive collisions ions-buffer gas into the RFQ chamber. Its optimum requires buffer gas pressures of a few Pa [22]. The vacuum system of an RFQC in general, and of SHIRaC in particular, has as a main challenge to keep the high vacuum at the injection and extraction parts whilst adding a high gas load into the RFQ chamber. It consists of separate sections connected by exchangeable apertures and the various pressures are obtained by differential pumping [13, 20].

The differential pumping system is very crucial for achieving the optimal cooling and transmission efficiency. The gas leakage through the entrance and the exit apertures cause the energy degradation of the beam and the related lost due to the scattering with atoms of the residual gas. These harmful effects can be reduced by careful design of the differential pumping system and the optimization of the placement of the injection and extraction electrodes with respect to the entrance and exit of the vacuum chamber where the RFQ is placed.

The main goal is to reach pressures less than of 0.01 Pa inside the injection and extraction parts and less than of 0.001 Pa in the rest of the beam-line. For this purpose, the injection and extraction plates have respectively, 6 and 4 mm of holes dimensions and 2 mm of size. Also, five turbo-molecular pumps with different pumping speeds are installed, figure 2.

Table 1 shows pressure measurements inside the different parts of SHIRaC beamline with various RFQ pressures (pressures into the RFQ chamber). We note, nonetheless, that the needed environment pressure is provided by this system.

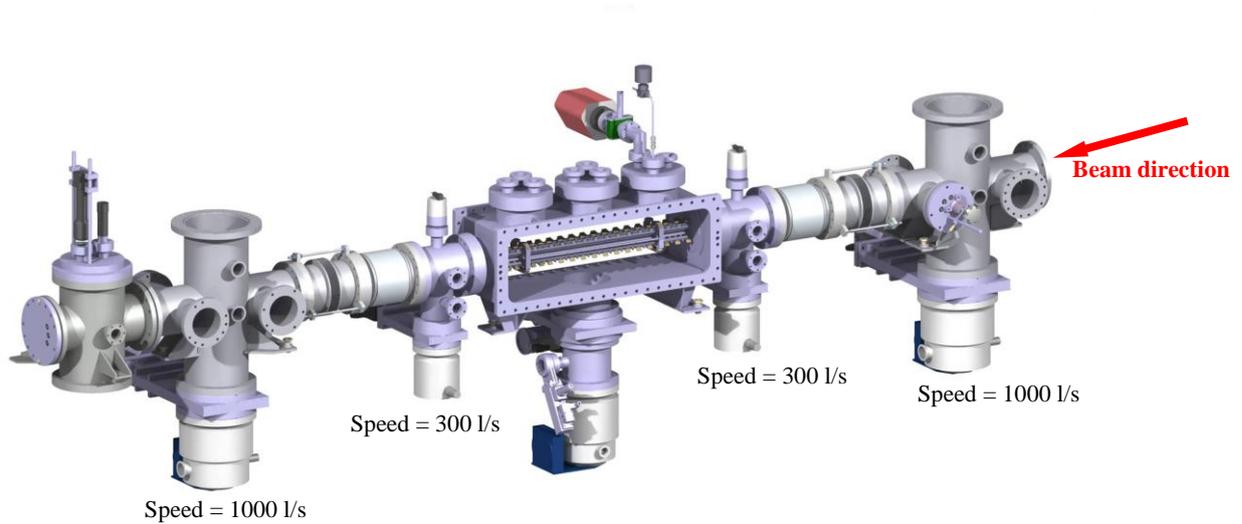

**Figure 2**: 3-D layout of the differential pumping vacuum system of SHIRaC.

| Gas rate (ml/mn) | Pressure in target region (Pa) | Pressure in extraction section (Pa) | Pressure in RFQ section (Pa) | Pressure in injection section (Pa) | Pressure in source section (Pa) |
|---|---|---|---|---|---|
| 30 | $2.0*10^{-3}$ | $9.1*10^{-3}$ | 3.1 | $2.1*10^{-2}$ | $3.5*10^{-3}$ |
| 25 | $1.5*10^{-3}$ | $6.6*10^{-3}$ | 2.5 | $1.6*10^{-2}$ | $2.8*10^{-3}$ |
| 20 | $1.3*10^{-3}$ | $5.4*10^{-3}$ | 2.1 | $1.2*10^{-2}$ | $2.3*10^{-3}$ |

**Table 1**: Pressure measurements inside the three section of SHIRaC beamline and at the source section as well as at the target region, between the extraction and the HRS.

## 2.3. Electronic system

In order to achieve the RF quadrupole field into the RFQ it was necessary to feed in two opposite phase sinusoidal RF voltages to the segment pairs. This resulted in all of the segments running together as four continuous quadrupole rods.

The RFQC requires applying an RF voltage, of some kV with some MHz [15, 16, 25], to the electrodes in presence of a buffer gas pressure of some Pa. To evaluate the sustainability of the necessary RF voltage in the operational pressure range, we developed a new RF system (figure 3). For the present work, a relatively strong RF confinement voltage is needed to overcome space-charge effects due to the high beam currents [22, 23, 24]. So, to achieve the high RF potentials across the trap electrodes at reasonable power levels, some form of resonating circuit has to be provided.

For the frequencies required for SHIRaC and the load levels presented, the only feasible resonant circuit is a lumped LC split series based on LC resonating circuit with transformer [26]. With such circuits, the coupling between the function generator and the LC circuit is done by a ferrite core transformer. Due to this material saturation [13, 26] the maximum allowable RF parameters are of about 1 kV and 1 MHz [25]. For the high-voltage operation required in the present work, to prevent failures caused by the saturation of transformer core, it is prudent to use a resonant LC circuit with air-inductive coupling, figure 3, where the coupling is done by a single turn injected into the hollow coils of the LC circuit. The LC circuits composed of a parallel air induction system (single coil) and a resonant LC circuit: two hollow coils (L1 and L2) and capacitor (CG 35-1000 pF), figure 3.

The RF system is a critical instrument of the RFQC because of the electric breakdown [22] and the ceramic insulator burning which can occur into the RFQ chamber for a few kV of RF amplitudes. To avoid the burning, the whole internal structure is done by the PEEK (PolyEther Ether Ketone), figure 4.

In order, to exploit the widest range of ion masses, a variable capacitor has been chosen.

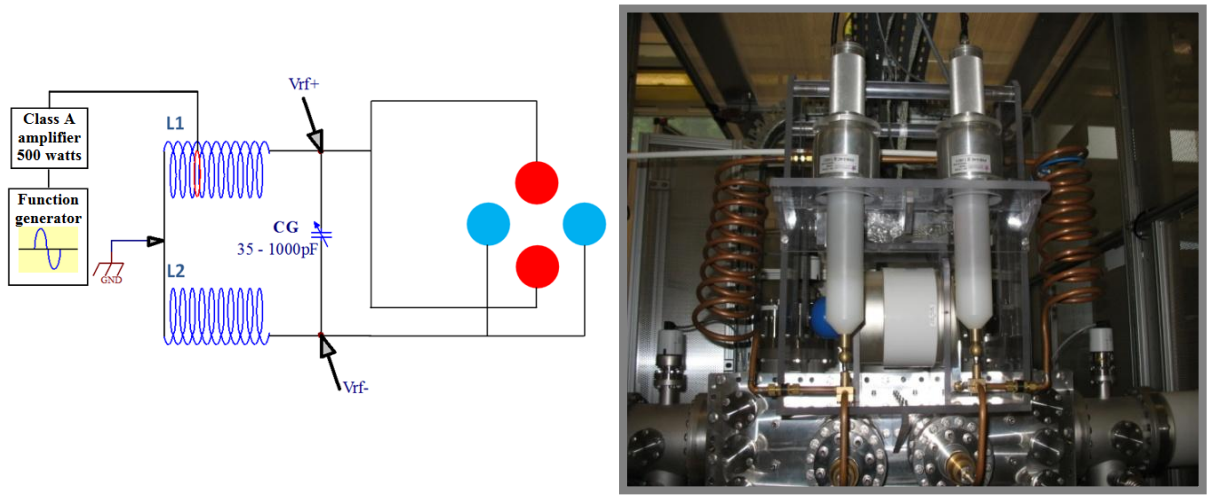

**Figure 3**: Design of the RF electronic system (left) and the real RF system of SHIRaC (right).

As we have seen previously, ions are cooled by collision on a light buffer-gas; subsequently they lose energy along the three dimensions. So, to allow them to pass through the quadrupole, it was necessary on top of the RF voltage, to couple a DC bias onto each segment pair, the bias level being determined by the segment pair's location along the axis. The most likely coupling method is a low pass filter with electronic component. But, in our case, RF amplitudes are too high to be blocked by conventional inductance. For that, the DC voltages were applied to the electrodes by threading separate DC leads through the interior of the RF resonating coils [15], figure 4, thus avoiding the complicated arrays of blocking inductors and shunting capacitors.

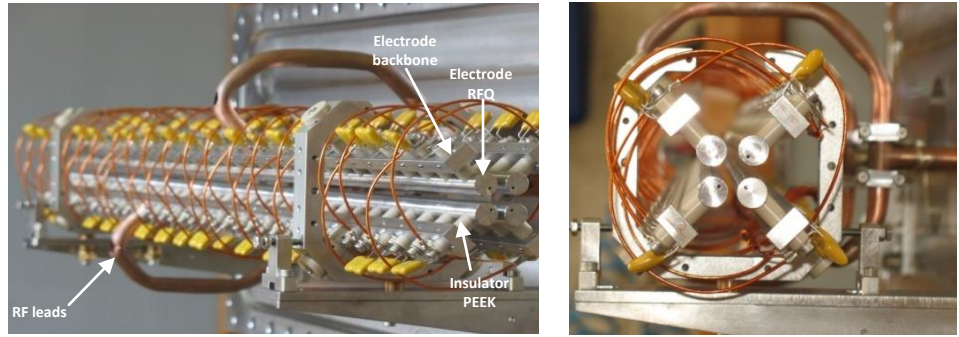

**Figure 4**: The voltages delivery scheme to the quadrupole electrodes of the RF supply and the power supplies for the DC axial guiding voltage.

Graphs of the results of RF voltages on the quadrupole's electrodes as a function of input voltage of the function generator, and with a LC resonant circuit of 8 turns per hollow coils, are shown in figure 5. Up to the allowable maximum input generator voltage 500 mV, it is seen absolutely that the RF circuit does not show any signs of being saturated and the output voltage (RF voltage) can even go up if any increasing of the input voltage occurs. This behavior shows the efficiency of such air-inductor coupling to produce higher RF voltage relative to the air-cores coupling. Thanks to the air-inductor coupling, the circuit can operate quite comfortably in delivering RF voltages up to 8 kV and 5 MHz on either side of the hollow coil. Furthermore, introducing helium into the RFQ chamber did not cause any noticeable problems until pressures approaching 5 Pa were reached.

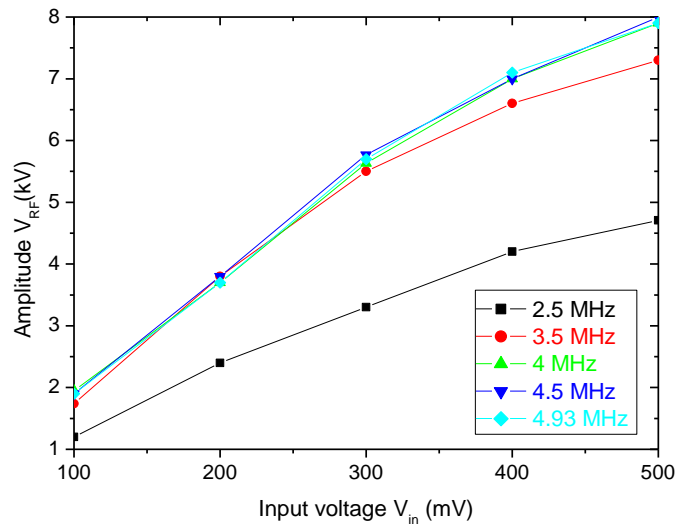

**Figure 5:** Plot of the RF voltage amplitudes as function of the input voltages, delivered by an amplifier of 500 W, for various frequencies set in the function generator and for a configuration of 8 turns per hollow coil.

The reduction of the cooling process degrading effect, due to the RF heating [22, 27], requires highest resonating frequencies $f_0$ [16]. For that, the hollow coils inductance had to be reduced. This was accomplished by testing the influence of the number of turns per coil on the resonating frequency, table 2. The RF performances improvements are, perceptibly, illustrated in this table. The resonating frequencies are enhanced and reach about 9 MHz. Moreover, the used RF circuit possesses a quality factors Q higher than the old coupling system where Q < 100 [26]. The high values of Q allow some ion beam impurities to be reduced during the cooling process because their stability depends strongly on the quality factor.

| Number of turns per coils | $f_0 (MHz)$ | $Q = \dfrac{f_0}{\Delta f_0}$ |
|---|---|---|
| 2 | 9 | 324 |
| 5 | 6,5 | 280 |
| 8 | 5 | 194 |

**Table 2**: Characteristics of the RF voltages delivered on the quadrupole's electrodes with various numbers of turn per coils.

## 3. Experimental Results

In order to confirm the performances of SHIRaC predicted by the simulations works [16] and enhance the cooled beam properties, we have completed several experimental studies focused on the specifications of this project. For that, we have used an IGS-4 ionization surface source which can provide $^{133}Cs^+$ ion beams with intensities up to 1 µA at energy up to 5 keV. Such beams have emittances of a few tens π mm mrad and energy spread of a few tens of eV. We will focus on the effects of the space charge, RF voltage and buffer gas pressure on the ion transmission and beam properties.

Tests of SHIRaC have been realized at LPC-Caen (France). A sketch of the experimental line is shown in figure 6. The RFQ chamber is preceded with the injection part connecting it to an ionization surface source. It is inside a differential pumping system and is followed by an extraction beam part equipped with beams diagnostics for current and beam measurements. The RF electronic system is mounted behind this chamber.

The experimental test has been realized with a RF frequency $f_0$ of 4.5 MHz. As the RF voltage amplitude $V_{RF}$ and the Mathieu parameter q are proportional [28, 29]:

$$V_{RF} = \frac{m r_0^2 \pi^2 f_0^2}{e} q$$

In this expression e, m and r0 are respectively, the elementary charge, the cesium ion mass and the quadrupole inner radius, it subsequently will be relevant to study the measurements dependencies on the parameter q than the RF amplitude. Under the same conditions stated above, this proportionality is of 6.9 for the cesium case.

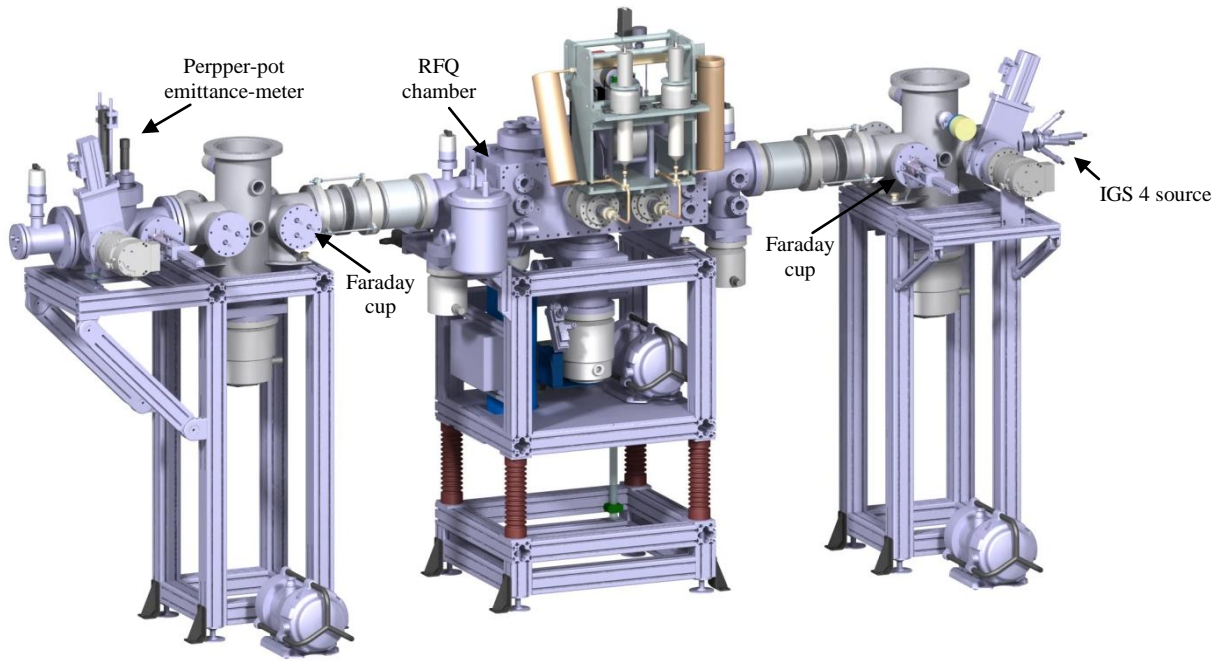

**Figure 6**: 3-D schematic of the planned test setup for SHIRaC prototype.

## 3.1. Transmission efficiency and operating parameters

To characterize SHIRaC, our first interest concerns to find the operating parameters via the transmission efficiency. We have performed these tests by studying the effects of the input ion energy, RFQ pressure and guiding DC voltage on the transmission. The transmission is the ratio of the beam current after the extraction section by the one at the injection section, measured by a cup Faraday.

The running parameters are the appropriate input energy (the energy the ions have when they enter the RF field), RFQ pressure and guiding DC voltage which can provide an optimum cooling, corresponding to the maximum ion transmission which depends on the simultaneous presence of these parameters. One way to determine the optimum value of each parameter is to fix two of them and vary the third one.

In fact, in the presence of a buffer gas the ions outgoing from the source with 5 keV of energy undergo a deceleration until reaching their input energy at the RFQ entrance. The ions with high input energy can ionize the gas atoms and thereafter degrade the cooled beam purity. But those with the appropriate energy undergo the cooling process along the RFQ. Their energy decreases progressively and they end up by stopping before they'd really reach the RFQ exit. To counter this stop and guide these ions up to the RFQ exit, without cooling degradation, a suitable DC guiding field is added.

The transmission dependencies on these parameters are presented in figure 7 and the appropriate values of input energy, buffer gas pressure and guiding DC voltage are respectively of 140 eV, 2.5 Pa and 16 V/m. In section 3.2, we will show that at such conditions the cooled beam possesses a high purity.

For more details we explain the behavior of the presented curves (at figure 7).

At figure 7-top, the transmission increase progressively with input ion energy to reach a plateau. Near to RFQ entrance, where no confinement field exists, undesirable ions-gas collisions can be done and result in ion losses. In order to minimize this phenomenon, the input energy should be as high as possible while preventing the gas ionization. So, the higher the input energy, the better transmission becomes. The least possible energy allowing the maximum transmission is around 140 eV. It is expected that the transmission increases with the buffer gas pressure [31], this is clearly illustrated in figure 7-middle. The observed decreasing beyond 2.5 Pa is due to the buffer gas diffusion towards the outside of the RFQ chamber. Finally, the ion transmission dependency on the guiding DC voltage (section 2.1) presented on figure 7-right shows a maximum transmission at 16 V/m. The transmission behavior explains a competition between the cooling effect, due to the gas pressure into the RFQ, and the ion accelerator effect due to the guiding field. The first effect is dominant for DC voltage below 16 V/m and the second one beyond.

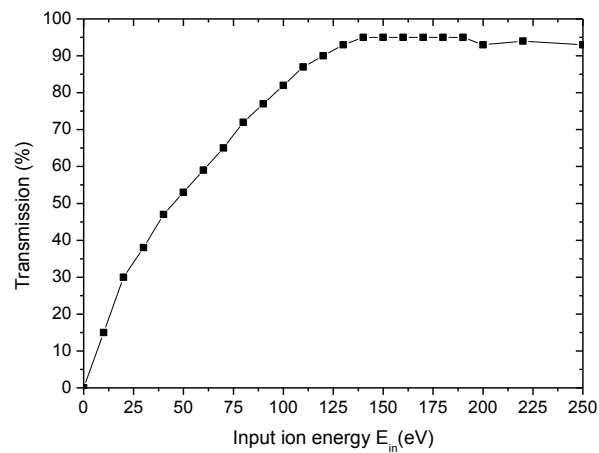

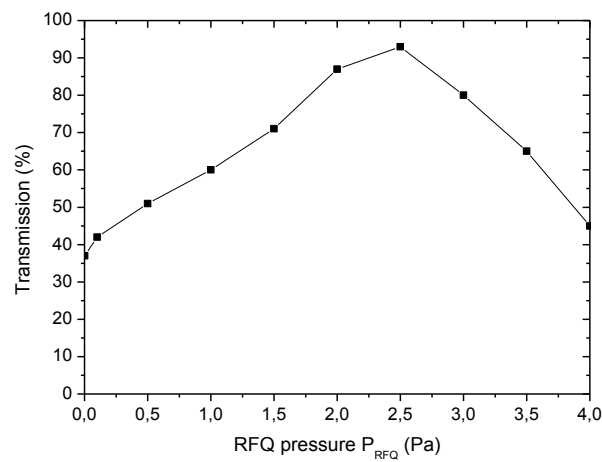

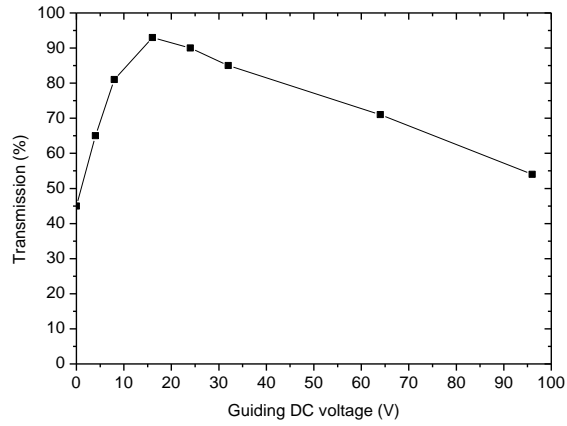

**Figure 7:** Operating parameters: Effect of the input ion energy ($E_{in}$) on the transmission for I= 50 nA of beam current, $P_{RFQ}$=2.5 Pa of RFQ pressure, $E_{DC}$=16 V/m of DC guiding field and q=0.4 (top). Buffer gas pressure effect on the transmission for I= 50 nA, $E_{DC}$=16 V/m, q=0.4 and $E_{in}$=140 eV (middle). Guiding DC voltage effect on the transmission for I= 50 nA, $P_{RFQ}$=2.5 Pa, $E_{in}$=140 eV and q=0.4 (down).

The simulation concluded that the influences of the space charge and RF voltage amplitude on the ion transmission are very important [16]. This induces us to study the dependencies of SHIRaC transmission on the beam currents and RF voltage amplitude.

The transmission behavior in dependence on the beam current was investigated in the figure 8. It gradually decreases with the beam current and remains above 70 % for current going up to 1 µA. This low loss of ions shows, on the one hand, the ability of the optics system to capture more than 93 % of incoming ions and on the other hand, it explains the efficiency of the high RF voltage amplitude to overcome the space charge effect and to confine such ions along the RFQ.

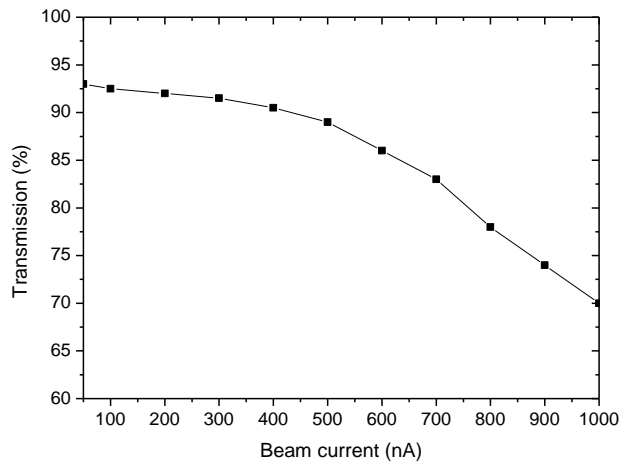

**Figure 8**: Space charge effect on the ion transmission: variation of the transmission versus beam current for a buffer gas pressure of 2.5 Pa.

The transmission dependence on the RF voltage i.e. the Mathieu parameter q is communally called the ion stability diagram. The figure below reveals this stability where ions can reach the RFQ exit for q between 0 and 0.9 [32]. The transmission as a function of q is shown in this figure. The transmission increases as q increases and reaches a maximum at q between 0.3 and 0.5 after which it decreases. For the buffer gas cooling beam guide, there is a threshold of potential at which the confining force is sufficient to overcome the defocusing effect by diffusion and space charge effect.

According to the reference [33], a widening phenomenon of the stability diagram can occur for high beam currents. For the present study, this phenomenon is not present for beam current going up to 1µA.

The relative difference between the red curve and the black curve shows once again that the optimum cooling is obtained with a RFQ pressure of 2.5 Pa.

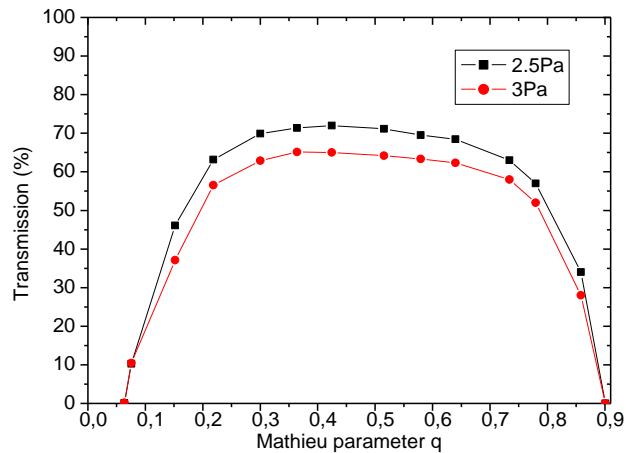

**Figure 9**: Stability diagram at 1 µA beam current: variation of the transmission as a function of the Mathieu parameter q for various buffer gas pressure.

## 3.2. Cooled beam purity

To detect the possible exchange of ionization between the cesium ions and the buffer gas or with possible impurities in the buffer gas [34], the ion beam purity was determined. To ascertain the presence of such impurities within the cooled beam, we have mounted a voltage switcher on the first electrode of the extraction lens, allowing to extract the cooled beam in mode bunch, and a MCP detector placed after the extraction section. The detector consists of an Einzel lens, an attenuator and a MCP screen, figure 10. As each ion reaches the MCP detector plane, its arrival time is binned appropriately, creating a time of flight histogram for each extracted ensemble that simulates the experimental ion signal recorded by the MCP detector. These signals imitate the temporal distribution of extracted cooled beam. As they are related to the ion mass, thus each peak of these signals will be associated to a set of ions of the same mass.

The measurement was assessed just under optimum cooling conditions and result is presented in figure 10-right. Ions other than cesium were not detected and this confirms that there is no ionization exchange between the cesium ion and the buffer gas or the impurities of the buffer gas.

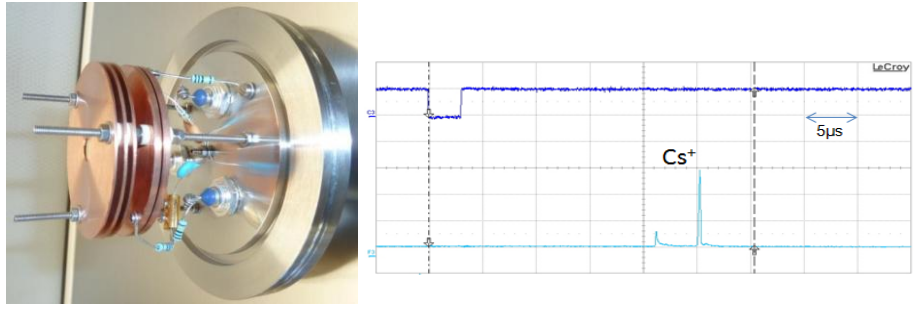

**Figure 10**: An overview of the MCP detector (Left) and cooled ions TOF Spectrum for 1 µA of beam current under the optimum cooling conditions (right).

### 3.3. Geometrical transverse emittance

For a determination of the geometric transverse emittance, one kind of measurements has been carried out with a commercial Pepper-pot emittance-meter [35] inserted perpendicularly to the beam direction and at the output of the beam line, figure 6.

The pepper-pot/multislit method is a common way to determine the emittance of ion beams. The pepper-pot mask consists of an even plate with a two dimensional array of holes in it. The part of the beam which passes the pepper-pot through its holes is separated into several spots with equal distance, thereafter hit a scintillating screen (MCP) located further downstream. The used pepper-pot has the following dimensions:

| Hole-hole distance (mm) | 1 ± 0.01 |
|---|---|
| Hole size (µm) | 100 ± 5 |
| Mask-MCP distance (mm) | 60 ± 0.2 |

**Table 3**: parameters and dimensions of the pepper-pot emittance meter.

The rms transverse emittance of a beam can be calculated from the position, the size, and the shape of these spots [36]. The geometric transverse emittance can then be deduced [13].
To deduce the equivalent emittance at 60 keV of beam energy, we can use the following relation [13]:

$$\varepsilon_{E_2} = \varepsilon_{E_1} \times \sqrt{\frac{E_1}{E_2}}$$

Where $\varepsilon_{E_1}$ and $\varepsilon_{E_2}$ are the beam emittances at beam energies of E1 and E2, respectively.
Thus, the emittance at 60 keV of beam energy is:

$$\varepsilon_{60\,keV} = \varepsilon_{5keV} \times \sqrt{\frac{5}{60}}$$

With this emittance meter is possible to follow the transverse emittance evolution. Figure 11 shows an emittance behaviour for different RF amplitudes (i.e. Mathieu parameter q) and for various beam currents.

These measurements are at once clearly sensitive to the RF voltage and the beam current. It is seen that, for all beam currents, the emittance first decreases to a minimum value as the RF voltage increases, and then slightly increases as the RF voltage is further increased. The rapid decrease of the curves explains the effect of the RF voltage to confine the ions and subsequently to enhance the beam properties. And then as the RF voltage is increased the emittance starts to saturate for q between 0.3-0.5, this corresponds to the maximum cooling power. For q more than 0.5, the emittance increases slightly. This degradation is due to the RF heating effect [22, 27] which arises for q > 0.5.

The emittance is also seen to increase in proportion to the ion beam intensity as expected from space charge considerations [16].

The optimum emittance values do not exceed 2.5 π mm mrad and are comparable to those for the previous Coolers, notwithstanding the beam intensities are often 10 times higher.

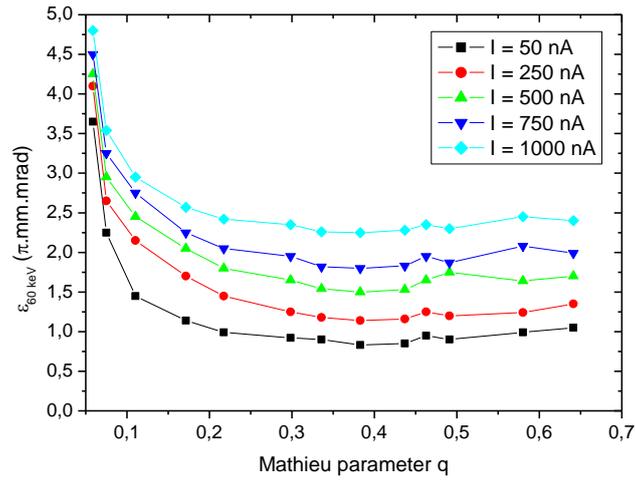

**Figure 11**: The geometric transverse emittance measurements versus Mathieu parameter q (i.e. RF voltage) for various beam currents under optimum cooling condition.

Measurement examples of emittances at lowest and highest beam currents are presented in figures 12 and 13, respectively. For 50 nA of beam current, the equivalent emittance at 60 keV is around 1 π mm mrad with an error does not exceed 10 %, figure 12. However, for 1 μA of beam current, the equivalent emittance at 60 keV is about 2.3 π mm mrad with an error around 30 %, figure 13. The increase of the error is due to the space charge effects which produce non linear distortions of beam emittance [37].

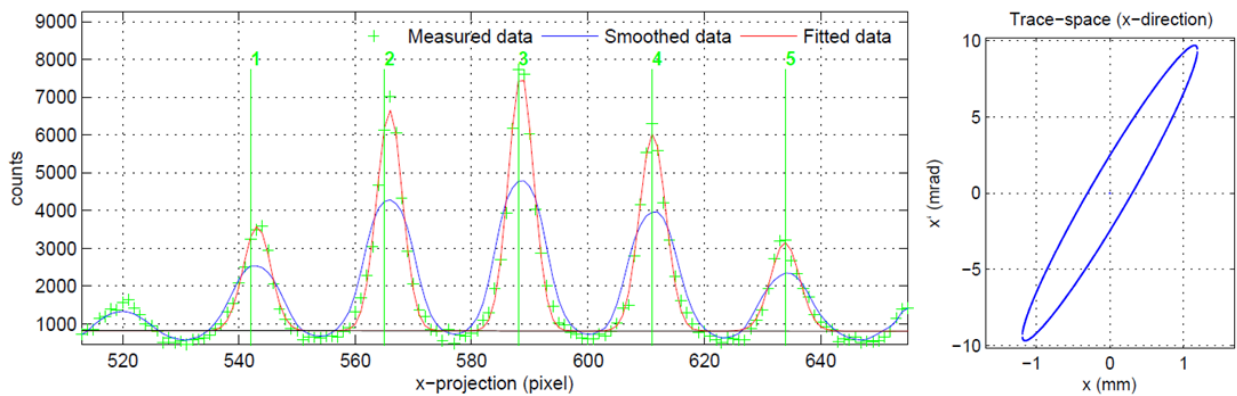

**Figure 12**: The cooled beam emittance measurement under the optimum cooling conditions for low beam current ( 50 nA) and with 0.4 of Mathieu parameter: the transverse distribution of cooled ions on the MCP screen and the multiple Gaussian fit of curves (left), the trace space x-direction and the geometrical emittance of $\varepsilon_{5keV}=1.9\pm 0.1$ π mm mrad at 5 keV which corresponds to 0.8 π mm mrad at 60 keV (right).

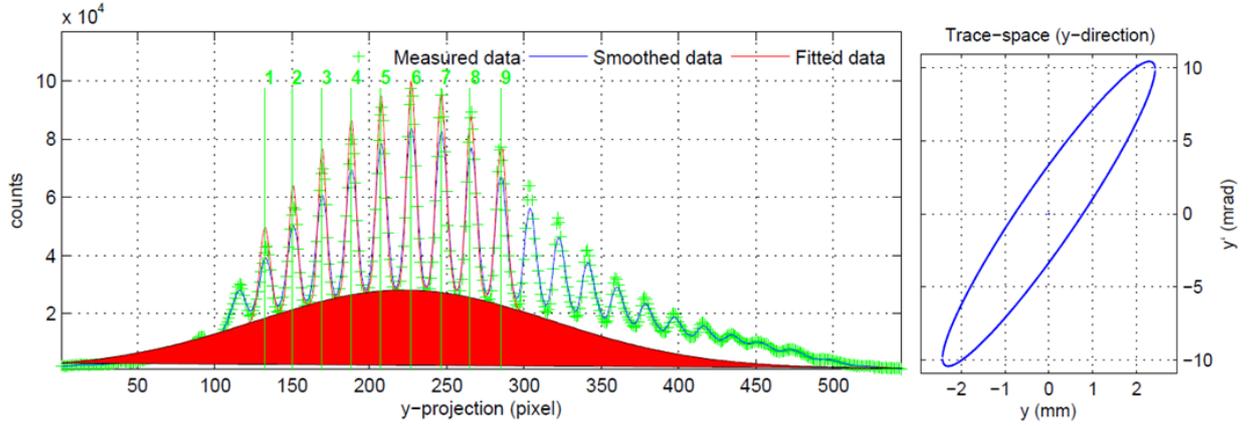

**Figure 13:** The cooled beam emittance measurement under the optimum cooling conditions for highest beam current ( 1 µA) and with 0.4 of Mathieu parameter: the transverse distribution of cooled ions on the MCP screen and the multiple Gaussian fit of curves (Left), the trace space x-direction and the geometrical emittance of $\varepsilon_{5keV}=8.1\pm 2.7$ π mm mrad at 5 keV which corresponds to about $2.3\pm0.7$ π mm mrad at 60 keV (right).

### 3.4. Longitudinal energy spread

Characterization of beam energy spread in a space-charge dominated beam is very important to understanding the cooling process of intense beams. In addition to the transmission and the emittance, the longitudinal energy spread is the third parameters to characterize the cooling performance of SHIRaC.

The measurement of the transmission variation as a function of the DC retarding voltage, applied on the last segment of the RFQ, is used to determine the longitudinal energy spread ΔE using the Full Width at Half Maximum (FWHM) of its derivative, figure 14.

The reduction of the energy spread is illustrated in the figure 14-right as regards its decrease with the increase of the buffer gas pressure $P_{RFQ}$. Above 2.5 Pa, an increase is seen and is explained by the degradation of the cooled beam properties by the buffer gas diffusion at the RFQ exit. Thus, the optimum cooling corresponds to a buffer gas pressure of 2.5 Pa.

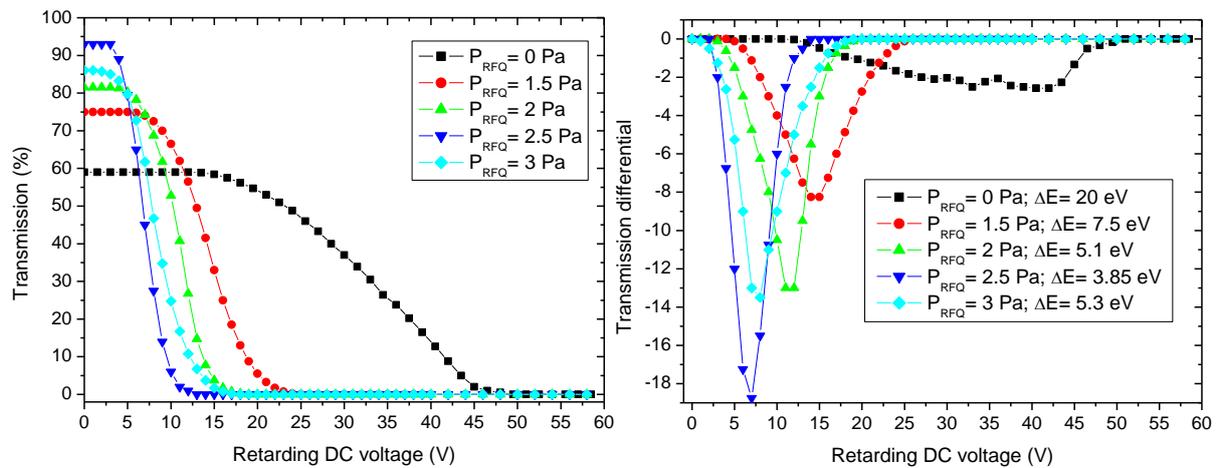

**Figure 14**: Measurements of the longitudinal energy spread with various buffer gas pressure $P_{RFQ}$ and for beam current of 50 nA: Variation of the transmission as a function of the retarding DC voltage (left) and variation of its differential versus the retarding DC voltage (right).

The simulation results presented in reference [16] have shown that the space charge has an important degrading effect on the energy spread. The below figure reports quantitative results of this phenomenon regarding the increase of the energy spread from 3.9 eV to about 6.5 eV for beam currents going from 50 nA up to 1 µA. These results are similar to those shown by the numerical simulations.

This degradation explains clearly the space charge effects, however the large energy spread arising at low beam current induces us to find the cause of this degradation.

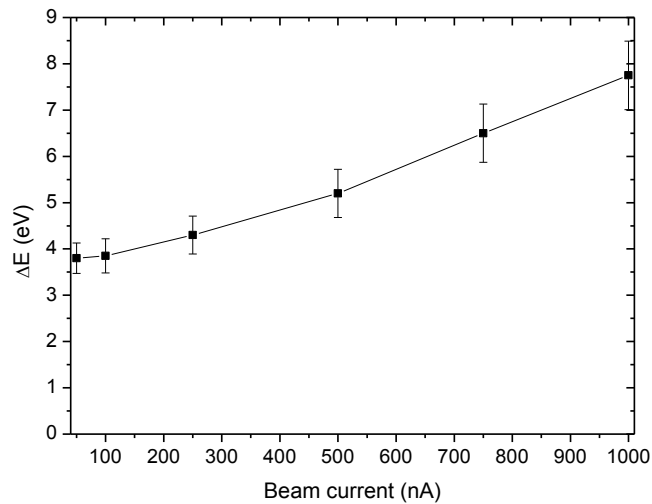

**Figure 15**: Space charge effect on the longitudinal energy spread: variation of the longitudinal energy spread with the beam current.

The main difference between the previous RFQCs and SHIRaC unwinds in the used RF voltage amplitude which is less than 1 kV in the first one versus more than 2.5 kV in the second one. This leads us to study the RF voltage effect on the energy spread. In figure 16 we present the variation of ΔE and its corresponding transmission as a function of the q parameter for beam current of 1 µA. As expected, for RF amplitudes less than 1 kV (q ≤ 0.15), the energy spread decreases with RF amplitude. However beyond 1 kV, an increase of the energy spread occurred.

This increase doesn't explain the effect of the RF amplitude on the cooling i.e. on the ΔE but explains the degradation of the cooled beam by the RF voltage at the RFQ exit region where the transversal RF field, created by the RF quadrupole, gives rise to an RF longitudinal field.

This field acts on the longitudinal velocities distribution of the cooled ions via its random characteristic of the RF voltage and thereafter results in a degradation of this distribution.

The minimum energy spread obtained with RF amplitude around 1 kV shows the competition between the cooling effect of the RF voltage and the degrading effect of the longitudinal RF field. At this minimum the ΔE is around 3.7 eV with a transmission of 38 %. For a transmission of at least of 60 % the longitudinal energy spread is around 5 eV.

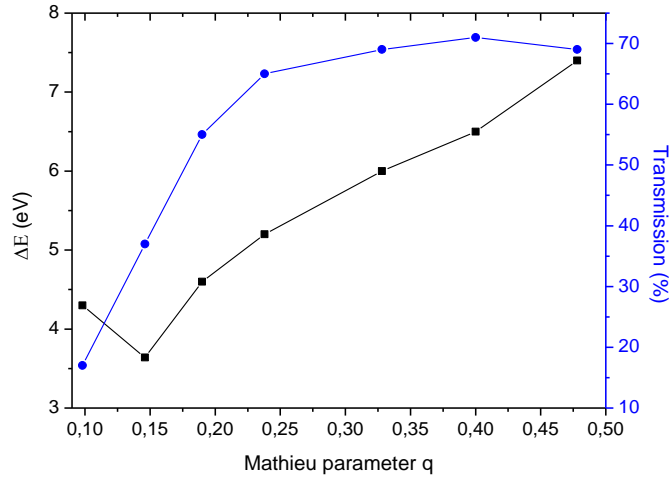

**Figure 16**: The RF voltage amplitude effect on the longitudinal energy spread: Variation of both the energy spread and its corresponding transmission as a function of the Mathieu parameter q.

## 4. Discussion

From the measured data and experience obtained in the above investigations the following conclusions can be drawn:

- The new RF system presented above allows reaching the highest RF parameters, up to 9 kV of amplitude and 9 MHz of frequencies without being limited by any breakdown problem. In such a condition the space charge effects can be adequately overcome.
- The optics system of SHIRaC and the appropriate voltages applied to the injection part electrodes show a good efficiency to capture highest beam currents and to transmit more than 70 % of incoming ions for currents going up to 1μA.
- For best cooling condition with SHIRaC, 140 eV of input energy, 2.5 Pa of RFQ pressure, Mathieu parameter q = 0.4 (RF amplitude of 2.8 kV and frequency of 4.5 MHz) and 16 V/m of guiding DC voltage are sufficient and they result in geometrical transverse emittance having low values not exceeding 2.5 π mm mrad with beam currents up to 1μA.
- The only disadvantage of this prototype emerges from the large values of the longitudinal energy spread. This is explained by the derivative effect of the RF field at the RFQ exit for amplitudes more than 1 kV.
- Both simulations [16] and experiments show close results, in terms of transmission and cooled beam parameters, for intensities never handled so far. For example, at the optimum cooling conditions of 1 μA beam, the simulation results are of 67 % of transmission, 2.5 π mm mrad of emittance and 5.9 eV of energy spread and their respective experimental results are of 70 %, 2.3±0.7 π mm mrad and 6.5±0.7 eV.

## 5. Conclusion

Before doing this SHIRaC prototype, the main challenge was to overcome the space charge stemming from the high beam currents. According to the numerical simulation presented in reference [16], the space charge effects emerge mainly at the RFQ exit region and they contribute to degrade both the transmission and the beam parameters. But, the large degradation of the energy spread with high beam current and even with low beam current could not be explained by these simulations. However, the present experimental studies show that the derivative of the RF confinement field, for RF amplitude more than 1 kV, can occur at the RFQ exit a longitudinal RF field which can be more degrading than the space charge. The latter is responsible for both transmission and emittance degradation but the second one acts mainly to enlarge the longitudinal energy spread.

To reduce the effects of both the space charge and the derivative of the RF field, a miniature RFQ will be installed at the RFQ exit [22]. This miniature RFQ aims to avoid the degradation of the cooled beam parameters by these effects and to guide these beams to a region where their energy is of a few tens eV and then they can resist to any degrading effects.

The longitudinal energy spread of the cooled beam is too high regarding the expected value which is around 1 eV. Therefore, an isobaric purification of this cooled beam by the HRS will become difficult and then a development of the extraction part will be necessary.

## Acknowledgment

We would like to thank Professor G.Ban for continuous encouragement and support.
The teams of electronic, vacuum system and mechanical design at L.P.C Caen (France) are gratefully acknowledged for their kind assistance in the development of the project.
We also thank Mouna yahyaoui for her English corrections.